\documentclass[11pt]{article}

\usepackage[utf8]{inputenc}
\usepackage{amsmath,amssymb}
\usepackage{graphicx}
\usepackage{dutchcal}
\usepackage{hyperref}
\usepackage{booktabs}
\usepackage{subcaption}
\usepackage{xcolor}

\title{Graphene Growth on Copper Substrate by LAMMPS Simulation}

\author{Lizhe Hong\thanks{Email: \texttt{lizhehong2-c@my.cityu.edu.hk}}\\[0.5em]
\small Department of Physics, City University of Hong Kong,\\
Kowloon Tong, Hong Kong SAR, China
}

\date{\today}

\begin{document}

\maketitle

\begin{abstract}
We learned the atomic deposition simulation of LAMMPS independently, referenced and optimized the modeling ideas of several papers, used the (1 1 1) crystalline surface of Cu atoms as a substrate, deposited C atoms produced by methane cleavage to obtain graphene flakes, and analyzed the deposition rate and deposition quality at three temperatures, obtaining conclusions consistent with the process flow.

We found that there were obvious problems in previous papers. After a certain period, the overall system pressure became excessively high, causing simulation crashes and preventing analysis of subsequent results. In addition, understanding of potential function selection was incomplete. Therefore, after correcting these issues, a simulation system with relatively stable pressure was constructed. In addition to the result analysis, a potential-function selection table is provided, with some parameters taken from prior experimental calculations and others obtained via DFT calculations.
\end{abstract}

\section{Introduction}

Thin film growth of 2D materials is a very interesting subject and the grown nanomaterials have some unique chemical-physical properties.

Graphene, as a common 2D material, is widely used in electronic devices, composite materials and biomedical applications. Currently, the main methods for preparing graphene include CVD, PVD, epitaxial growth method and electrochemical exfoliation method.

In this thesis, the process will be investigated through the idea of simulation. Reproducing the whole process on a macroscopic scale requires the use of multiphysics field simulation software (e.g., COMSOL FDTD, etc.), which has long computation time and complex modeling steps, making it difficult to study the statistical and micromechanical properties of the process. Therefore, in this study, as a tutorial nature exploration, the molecular dynamics software LAMMPS will be used to simulate the deposition process of graphene thin films and to correct some modeling as well as analytical problems in previous studies.

\section{Computational Method}
\subsection{Growth Environment Setup}

This simulation is modified based on the study of Zhang et al. The size of the selected substrate area as well as the computational time are limited due to the computational conditions.

Selection of the simulation box part: the size of this simulation was selected as a box with a volume about four times the size of the deposition system. Because the thermal motion of the substrate in this process is more intense, all boundary conditions are non-periodic.

Atomic selection: Cu atom atomic mass: 63.5 g/mol radius 1.28 \AA. 

Dissociative carbon atom (C1) atomic mass: 12.0 g/mol radius: 0.77 \AA. 

Deposition of carbon atoms (C2) atomic mass: 12.0 g/mol radius: 0.77 \AA. 

Substrate Selection: The lattice constant of the Cu atom is selected as 3.615 \AA, to set up an initial substrate, leaving the free space is large to prevent the system pressure is too large, the use of orient command on the substrate steering, so that the C atoms deposited in the Cu (1 1 1) crystal surface, which makes C atomic energy is approximately stabilized in the gaps of Cu atoms.The modeling is completed using the cg conjugate gradient minimization way to optimize the Cu atoms substrate to avoid excessive stress.

Generation layer and deposition layer selection: A generation layer of C1 atoms is defined above the original crystal plane as the dissociated part of the C atoms, which are renewed as C2 atoms when they fall close to the Cu crystal plane.

Reflective wall: A reflective wall for C atoms is set up in the outer shell of the Cu substrate to prevent C atoms from falling to the side or bottom of the Cu substrate.

\subsection{Potential function}
\begin{table}[!ht]
    \centering
    \caption{Force Field Parameters}
    \footnotesize
    \renewcommand{\arraystretch}{1.0}
    \resizebox{\textwidth}{!}{%
    \begin{tabular}{cccccccccc}
    \toprule
    \textbf{Type} & \textbf{LJ $\epsilon$} & \textbf{LJ $\sigma$} & \textbf{LJ $r_c$} & \textbf{Morse $D_e$} & \textbf{Morse $\alpha$} & \textbf{Morse $r_e$} & \textbf{Morse $r_c$} & \textbf{EAM} & \textbf{AIREBO} \\
    \midrule
    Cu-Cu & * & * & * & 0.3429 & 1.3588 & 2.860 & 8.37387 & Open & Close \\
    Cu-C1 & * & * & * & 0.1 & 1.70 & 2.22 & 6.5 & Close & Close \\
    Cu-C2 & * & * & * & 0.1 & 1.70 & 2.22 & 6.5 & Close & Close \\
    C1-C1 & 0.01276 & 3.81 & 9.525 & * & * & * & * & Close & Close \\
    C1-C2 & * & * & * & * & * & * & * & Close & Open \\
    C2-C2 & * & * & * & * & * & * & * & Close & Open \\
    \bottomrule
    \end{tabular}%
    }
    \label{tab:force_field_parameters}
\end{table}

In this study, different potential functions are selected for different atoms and between the two, and for the base Cu atoms between them, the overlay command, which is a superposition potential, is used for their calculation.

LJ potential: as the most common type of potential energy in LAMMPS simulation, also called hard-sphere potential, generally used to describe the interaction force between two non-bonded atoms, the standard formula is as follows:

\begin{equation}
U(r) = 4\epsilon \left[ \left( \frac{\sigma}{r} \right)^{12} - \left( \frac{\sigma}{r} \right)^{6} \right]
\end{equation}
\hspace*{1em}\( U(r) \): The potential energy between two particles. \\
\hspace*{1em}\( r \): The distance between the particle centers. \\
\hspace*{1em}\( \epsilon \): The depth of the potential well, indicating interaction strength. \\
\hspace*{1em}\( \sigma \): The characteristic distance where the potential is zero. \\
\hspace*{1em}\( 2^{1/6} \sigma \): The equilibrium distance where the potential is minimized. \\
\hspace*{1em}Cutoff distance: A distance beyond which the interaction is neglected in simulations.

Morse potential: as a more stable interaction potential in LAMMPS simulation, generally selected in the metal-C atomic system, the formula is as follows: 
\begin{equation}
U(r) = D_e \left( 1 - e^{-a(r - r_e)} \right)^2
\end{equation}
\hspace*{1em}\( D_e \): The dissociation energy, the well depth, which represents the energy required to separate the two particles. \\
\hspace*{1em}\( r_e \): The equilibrium distance, where the potential energy is at a minimum. \\
\hspace*{1em}\( a \): The parameter that controls the steepness of the potential well. A larger \( a \) results in a steeper potential.
Eam potential: as a composite potential, the principle is More complex, generally describes the interactions of atoms between metals.

Airebo potential energy: as the optimal solution of bonding between atoms of the multi-system composite potential energy, the equilibrium distance can be approximated as the bonding distance, containing a variety of potential energy, can be obtained by DFT calculation.

Abandoned potentials: tersoff potentials are used to model covalent bonding between atoms and are suitable for the deposition of complexes of inorganic systems, e.g., Si-C systems, but should be abandoned if the substrate is a metal.

All the potential parameters used are shown in Table~\ref{tab:force_field_parameters}.

The interaction energy between C1 and C2 atoms should be kept consistent with that between two C2 atoms, otherwise the system will undergo a sudden change in the process of deposition renewal resulting in an excessive transient velocity, while the temperature is the macroscopic result of the particle motion, thus leading to the failure of NVT temperature control and excessive pressure or high temperature of the whole system.
\subsection{Thermodynamic setup and Particle update method}

The heat bath system is divided into three parts: 

Gas region: blank part, C1 atom part 

Active solid region: C2 atom part, upper Cu atom part 

Solid part: lower Cu atom part 

\begin{figure}[!ht]
    \centering
    \includegraphics[width=0.75\linewidth]{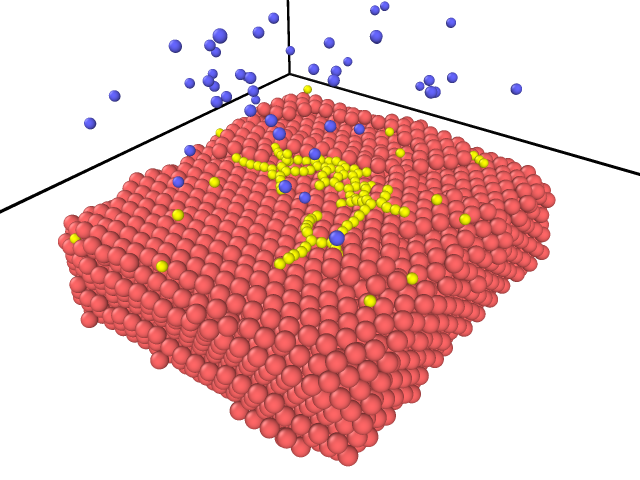}
    \caption{Structure of the simulated system: red atoms are Cu atoms, blue atoms are C1 atoms, yellow atoms are C2 atoms.}
    \label{fig:structure}
\end{figure}

The entire system after modeling is shown in Fig.~\ref{fig:structure}.

For the C1 atom and C2 atom part, the NVT system needs to be adopted for the gradual heating, the starting temperature is set to be 300 K (close to the room temperature), and the heating rate is set to be 0.01 K for each step, and time step is set to be 1 fs.

Cu atoms need to be calculated after cg energy minimization, otherwise it will cause crystal defects as well as excessive pressure, and it is difficult to form a slight depression on the surface to provide a deposition area. 

For the Cu atoms in the upper layer, it is necessary to carry out a Langevin thermal bath, and it can not be included in the C2 atoms, otherwise it will result in the atom perturbation speed is too fast and directly passes through the reflective wall.

For the fixed Cu atoms, the fix command is used to make the force constant equal to 0, fixing them in the system.

\section{Results and Discussion}

Regarding this simulation, we combined the melting point of Cu atoms (1357.6 K) and the temperature range of the system, the target temperatures selected were 900 K, 1100 K, 1300 K, and the simulation time was 500 ps. The part of the experimental results we analyzed quantitatively by combining the intuitive deposition conformation of the C2 on the surface of the Cu atoms' substrate and its radius of gyration with the change of time, so as to determine the good or bad growth of graphene. The growth efficiency of the system can also be determined by analyzing the number of stable structures and their ratio to the number of atoms deposited in the system at 500 ps. In addition, the deposition rate of C atoms is 5 Atoms × 0.1 \AA/ps, and we also plotted the deposition number-time curve to determine the most suitable temperature range for the growth of graphene underneath the Cu-atom base under the industrial system by the combination of the deposition efficiency as well as the growth quality. The radius of gyration was calculated as follows:
\begin{equation}
    R = \sqrt{\frac{1}{N} \sum_{i=1}^{N} r_i^2}
\end{equation}

\begin{figure}[!ht]
    \centering
    \begin{subfigure}[b]{0.3\textwidth}
        \centering
        \includegraphics[width=\linewidth]{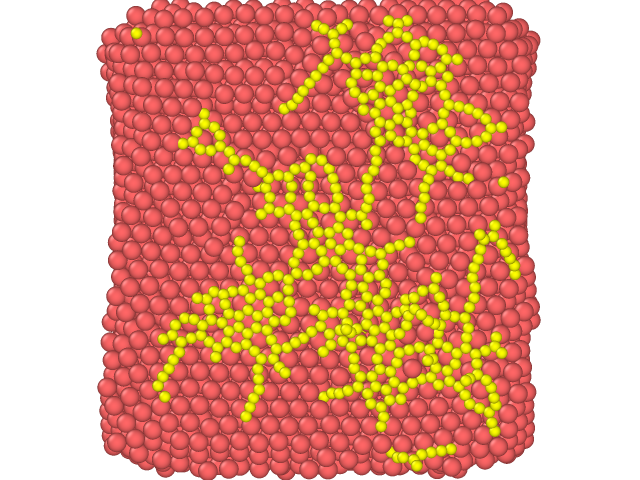}
        \caption{}
    \end{subfigure}
    \hfill
    \begin{subfigure}[b]{0.3\textwidth}
        \centering
        \includegraphics[width=\linewidth]{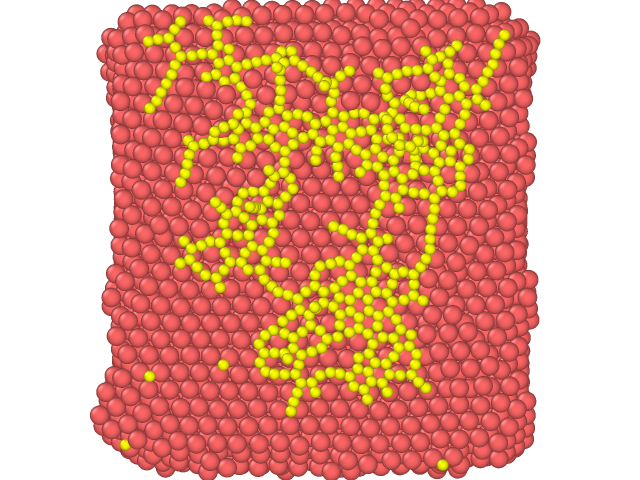}
        \caption{}
    \end{subfigure}
    \hfill
    \begin{subfigure}[b]{0.3\textwidth}
        \centering
        \includegraphics[width=\linewidth]{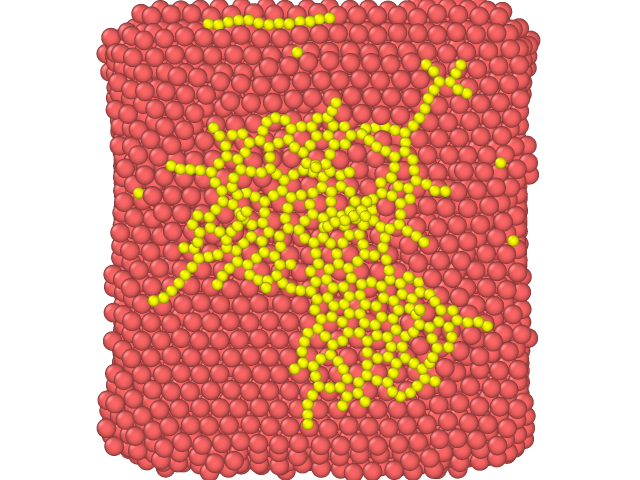}
        \caption{}
    \end{subfigure}
    
    \begin{subfigure}[b]{0.45\textwidth}
        \centering
        \includegraphics[width=\linewidth]{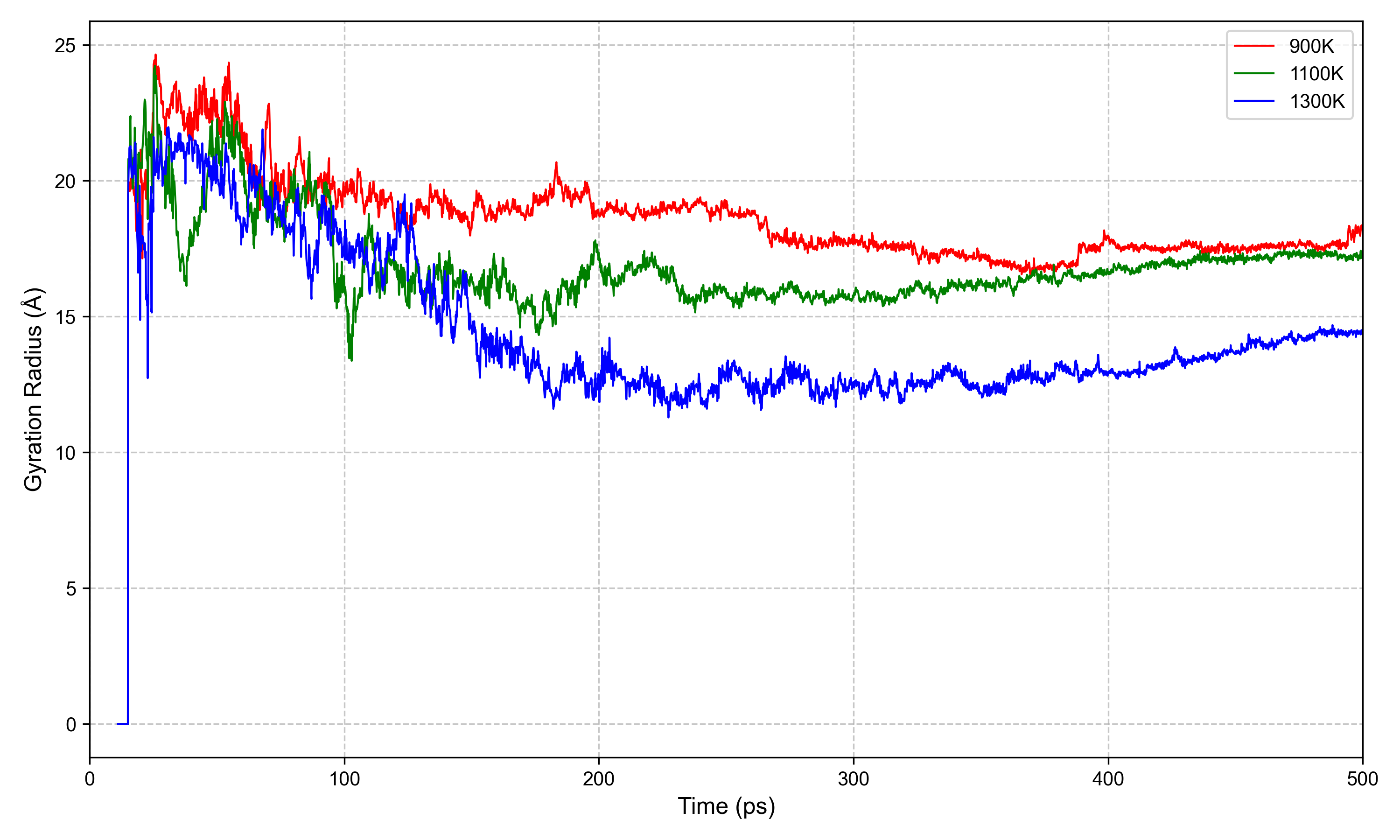}
        \caption{}
    \end{subfigure}
    \hfill
    \begin{subfigure}[b]{0.45\textwidth}
        \centering
        \includegraphics[width=\linewidth]{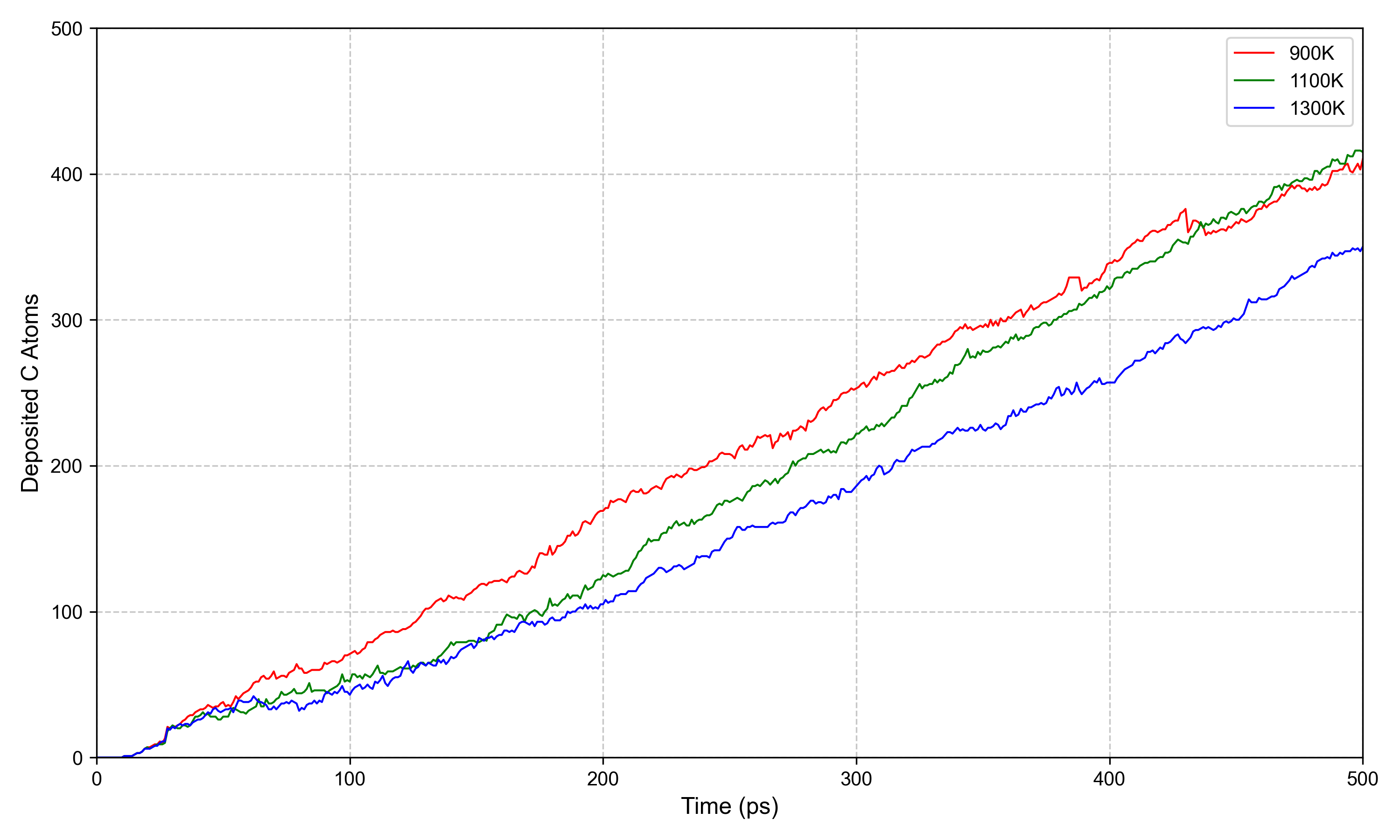}
        \caption{}
    \end{subfigure}
    
    \begin{subfigure}[b]{0.45\textwidth}
        \centering
        \includegraphics[width=\linewidth]{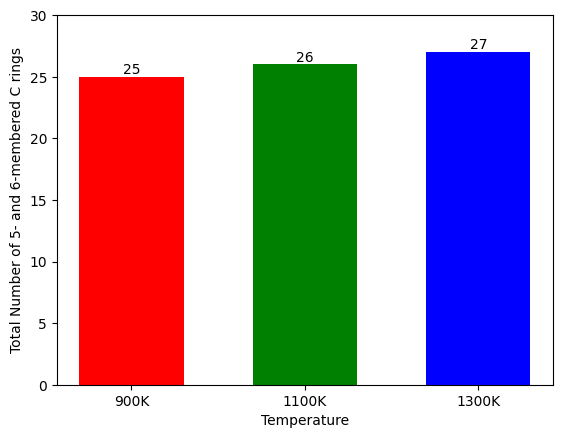}
        \caption{}
    \end{subfigure}
    \hfill
    \begin{subfigure}[b]{0.45\textwidth}
        \centering
        \includegraphics[width=\linewidth]{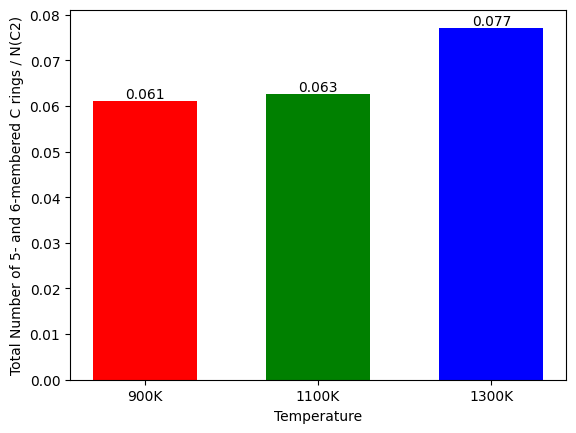}
        \caption{}
    \end{subfigure}
    
    \caption{Top views of the overall system at different temperatures and other relevant distributions. (a) 900 K, (b) 1100 K, (c) 1300 K. (d) time-swing radius distribution images. (e) time-deposited atom number distribution images. (f) In t=500 ps, number of five-membered rings. (g) ratio of five-membered rings to C atoms in the total deposited region.}
    \label{fig:images}
\end{figure}
    
Where the physical quantities are defined as follows:
    
\hspace*{1em}\( R \): The three-dimensional radius of gyration, representing the root-mean-square distance from the center of mass.

\hspace*{1em}\( N \): The total number of particles in the considered cluster.

\hspace*{1em}\( r_i \): The distance of the \( i \)-th particle from the center of mass.
\section{Radius of gyration and visual representation of C2}
\subsection{Explanation based on perturbation theory}

The nucleation process of graphene on Cu is influenced by the temperature and the density of C atoms available for deposition. At lower temperatures, the graphene islands remain small, and the nucleation density is relatively low. As the temperature increases, the nucleation density increases, and larger graphene islands begin to form. The temperature-dependent evolution of the nucleation process is shown in Fig. 2(a)-(c).

\subsection{Graphene Growth and Layer Formation}
The top view of C-atom deposition at different temperatures is shown in Fig. 2(a)-(c), from which it can be seen that the continuity of the growth of graphene gradually strengthens with the increase of temperature at 900 K-1300 K. However, it is not qualitatively represented, so we extracted the coordinates of the C-atoms of the deposition layer and plotted the Fig. 2(d) time-radius-of-swing distribution curve , which was obtained as follows: as the deposition progresses gradually, in most of the time, the radius of gyration of graphene in this temperature range decreases with the increase of temperature, and there is an obvious abrupt change from 1100 K to 1300 K, which implies not only the conformational densification, but also a sudden rise in the quality of graphene growth.
\subsection{Rate of adsorption of C atoms}
In addition, the adsorption rate of C atoms is also an important factor to measure the growth of graphene, the higher adsorption efficiency of C atoms means the faster growth of graphene, i.e., the increase of the yield, so we extracted the time-deposition quantity curve of C atom deposition and obtained Fig. 2(e).

From the figure, we can see that the overall image shows a gradual decrease in the deposition rate of atoms with increasing temperature for most of the time, which is not explained in detail or theoretically conjectured in many literatures. Therefore we combined the observation of the simulation animation with the visualization of the cause used to explain this phenomenon, as shown in Fig. 3(a):

With this figure we can recognize that larger perturbations in temperature tend to make incompletely deposited C atoms fly out, which is why the higher the temperature the slower the deposition rate is during the incomplete growth of graphene

Fig. 2(f) and (g) represent the sum of the number of 5-membered rings and 6-membered rings in graphene stretching at different temperatures under 500 ps, and their ratio to the total number of deposited C atoms. It is not difficult to find that at different temperatures, the number of 5-membered rings and 6-membered rings, that is, the part of graphene that can be regarded as a stable structure, has steadily increased, but the difference is not large. However, by dividing the quality of graphene growth by the deposition rate, we found that the quality of graphene growth has a significant increase from 1100 K to 1300 K, which also explains why graphene growth needs to be selected at 1300 K. One reason is that the utilization rate of deposited C atoms is higher, and the other is that the arrangement of stable rings is more compact, and graphene is not easy to break.

Previous studies have summarized experimental data but lacked theoretical explanations for the system. Therefore, we made a hypothesis and used Fig.~\ref{fig:FIG3}(b) for analysis:

\begin{figure}[!ht]
    \centering
    \begin{subfigure}[b]{0.45\textwidth}
        \centering
        \includegraphics[width=\linewidth]{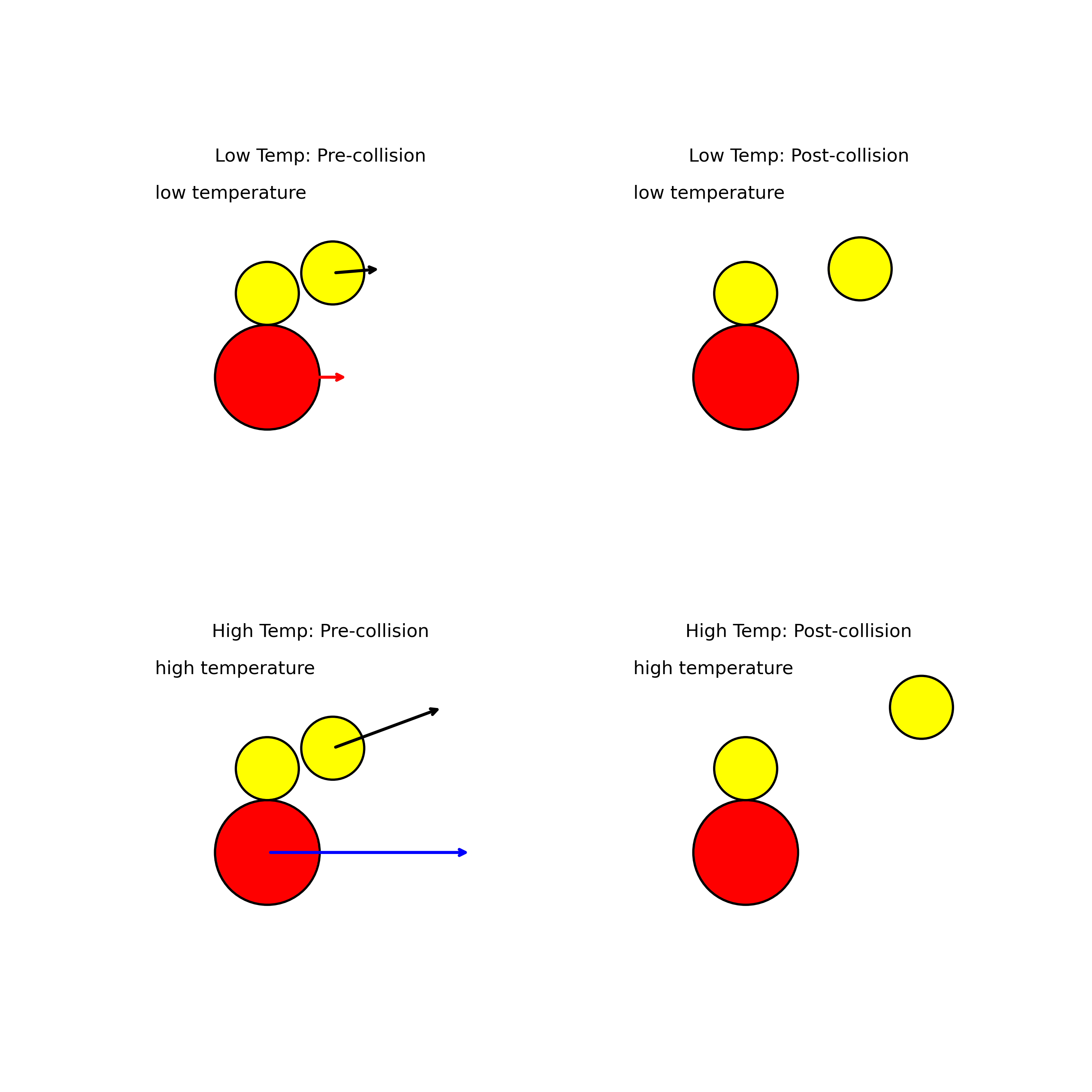}
        \caption{}
    \end{subfigure}
    \hfill
    \begin{subfigure}[b]{0.45\textwidth}
        \centering
        \includegraphics[width=\linewidth]{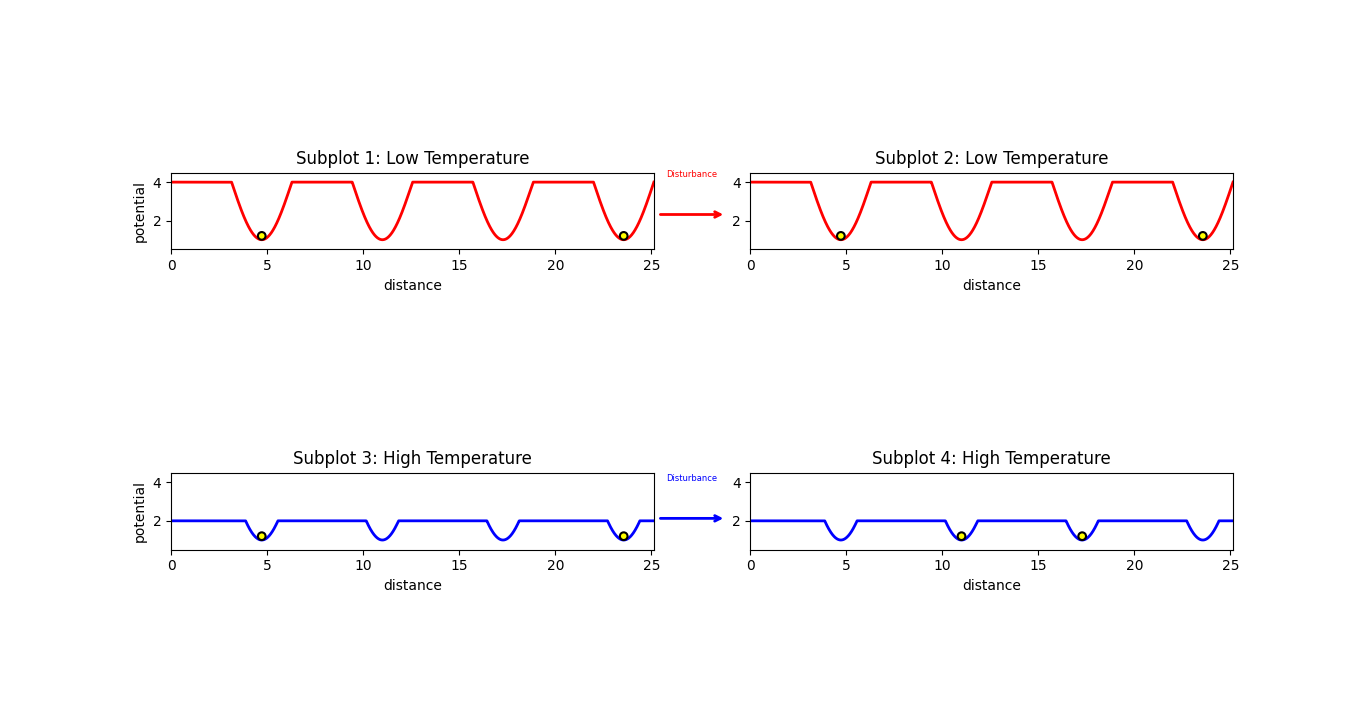}
        \caption{}
    \end{subfigure}
    \caption{(a) Because of the Langevin heat bath, the high-temperature Cu atoms pull the already adsorbed C atoms to hit the C atoms that are not adsorbed with the Cu atoms, thus causing the C atoms to fly out, where the red arrows represent the amplitude of the movement of the Cu atoms at lower temperatures, and the blue arrows represent the amplitude of the movement of the Cu atoms at higher temperatures. (b) Potential energy curve image, the horizontal axis represents the distance between C atoms, and the vertical axis represents the potential energy, showing the reason why C atoms grow densely at high temperatures.}
    \label{fig:FIG3}
\end{figure}

The disturbance of Cu atoms in the Langevin heat bath increases with the increase of temperature. After C atoms attach to Cu atoms, they will randomly walk due to the disturbance of Cu atoms. Because of the airebo potential energy, C-C atoms are mutually attracted to each other in most non-bonding distances. The increase in disturbance is equivalent to the decrease in energy potential wells, making it easier for C atoms farther away to escape from the gap of Cu and fall into adjacent potential wells, increasing the probability of collision. This conjecture is based on observation of the animation and can be used to explain the growth problem of this system.

Therefore, we believe that the most suitable temperature for C atoms to grow into graphene should be around 1300 K. Although its deposition rate is slightly lower, the stable utilization rate of C is higher, and the grown graphene is relatively stable. It is not easy to break and fall off from the substrate during the process, which is consistent with the previous simulation.

\section{Conclusion}
CVD cracking methane was used in this simulation, which focuses on analyzing the adsorption efficiency and growth quality of graphene growth and some theoretical speculations of its growth process by focusing on a different study from the previous ones.C atoms will fall into the gap of Cu atoms to form a stabilized structure of a 5-6-membered ring, and the probability of the emergence of a stabilized structure is significantly increased at around 1300 K. Therefore, the actual process of the around 1300 K to explore the appropriate temperature and appropriate pressure control, so as to obtain continuous graphene in sheets.
\section{Acknowledgments}
No prior experience related to atomic deposition was held, and few tutorials currently exist that successfully simulate 2D-material growth in papers or forums. The simulation was initiated from personal interest and one week of self-study with minimal prior knowledge. The significance of the various potential functions, as well as the selection and modification of the force fields, was thoroughly understood. Acknowledgment is given to Yu Congbo and Xiao Ma for providing LAMMPS atomic deposition tutorials and modeling ideas. After the course, the relevant code will be made publicly available on GitHub at \href{https://github.com/Showleader}{https://github.com/Showleader} for community learning and exchange.

\end{document}